\documentclass[letter,longauth]{aa}  
\usepackage{natbib} 
\usepackage{txfonts}
\usepackage{graphicx}
\usepackage{amssymb}

\newcommand{\num} {$\nu_{\rm max}$\xspace}

\newcommand{\dnu} {$\Delta\nu$\xspace}

\newcommand{\Gaia} {\textit{Gaia}\xspace}

\newcommand{\GaiaDR} {\textit{Gaia}\,DR3\xspace}

\newcommand{\apol} {\texttt{apollinaire}\xspace}

\newcommand{\Kepler} {\textit{Kepler}\xspace}

\newcommand{\KIC}{KIC\,9693187\xspace}

\def\dnu{$\Delta\nu$\xspace}
\def\dn1{$\delta\nu_{01}$\xspace}
\def\dn2{$\delta\nu_{02}$\xspace}

\usepackage[colorlinks=true,
                        linkcolor=cyan,
                        citecolor=cyan, 
                        filecolor=cyan, 
                        urlcolor=cyan]{hyperref}

\begin{document} 

\title{
Dynamical mass of a solar-like oscillator at the main sequence turnoff from \Gaia astrometry \& ground-based spectroscopy}

\titlerunning{\KIC $-$ Dynamical Masses for a Main-Sequence Solar-like Oscillators}
\authorrunning{Beck\,et\,al.}

\author{P.\,G.\,Beck\inst{\ref{inst:IAC},\ref{inst:ULL}} 
\and T.\,Masseron\inst{\ref{inst:IAC},\ref{inst:ULL}}
\and K.\,Pavlovski\inst{\ref{inst:Zagreb}}
\and D.\,Godoy-Rivera\inst{\ref{inst:IAC},\ref{inst:ULL}}
\and S.\,Mathur\inst{\ref{inst:IAC},\ref{inst:ULL}}%
\and D.\,H.\,Grossmann\inst{\ref{inst:IAC},\ref{inst:ULL}}\and
A.\,Hamy\inst{\ref{inst:CEA},\ref{inst:Ecole}}\and
D.\,B.~Palakkatharappil\inst{\ref{inst:CEA}} 
\and E. Panetier\inst{\ref{inst:CEA}} 
\and R.\,A.\,García\inst{\ref{inst:CEA}}
\and J.\,Merc\inst{\ref{inst:Charles},\ref{inst:IAC}}
\and Y. Lu$^{\ref{inst:Exeter}}$ 
\and I.\,Amestoy\inst{\ref{inst:TLU}}
\and H.\,J.\,Deeg\inst{\ref{inst:IAC},\ref{inst:ULL}}
}

\institute{Instit.\,de\,Astrof\'{\i}sica\,de\,Canarias,\,E38205\,La\,Laguna,\,Tenerife,\,Spain \label{inst:IAC}
\and Departamento\,de\,Astrof\'{\i}sica,\,Univ.\,de\,La\,Laguna,\,Tenerife,\,Spain \label{inst:ULL}
\and  Department of Physics, Faculty of Science, Univ. of Zagreb, Croatia \label{inst:Zagreb}
\and Universit\'e Paris-Saclay, Universit\'e Paris Cit\'e, CEA, CNRS, AIM, 91191, Gif-sur-Yvette, France \label{inst:CEA}
\and École\,CentraleSupélec,\,Univ.\,Paris-Saclay,\,Gif-sur-Yvette,\,FRA\label{inst:Ecole}
\and Astronomical Institute, Faculty of Mathematics and Physics, Charles University, V Holešovičkách 2, 180 00 Prague, Czechia\label{inst:Charles}
\and Astrophysics Group, School of Physics and Astronomy, University of Exeter, Stocker Road, Exeter EX4 4QL, UK\label{inst:Exeter}
\and IRAP, Université de Toulouse, CNRS, CNES, UPS, 14 avenue Edouard Belin, 31400 Toulouse, France \label{inst:TLU}
}

\date{Submitted: September 27, 2025; Accepted: January 17, 2026}

\abstract{Asteroseismology is widely used for the precise mass determination of solar-like oscillating stars, based on individual frequency modeling or homological scaling relations. However, these methods have not been dynamically validated on the main sequence (MS) due to the absence of eclipsing double-lined binary system (SB2) as benchmark objects. By providing the orbital inclination, astrometric binary systems from ESA \GaiaDR offer an abundant alternative for eclipsing systems. 
We present \KIC as the first SB2 hosting a solar-like oscillating post-MS star with dynamical masses. By combining \Gaia astrometry with spectroscopic data obtained with the Las Cumbres Observatory network (LCO),  we found $M_1^\mathrm{dyn}$\,=\,0.99\,$\pm$0.05$M_\odot$ and $M_2^\mathrm{dyn}$\,=\,0.89\,$\pm$0.04$M_\odot$ for the primary and secondary, respectively. The asteroseismic parameters were extracted from photometry of the NASA \Kepler satellite.  The mass from individual frequency modeling is $M_1^\mathrm{IF}$\,=\,0.92\,$\pm$0.01$M_\odot$. Taking into account the systematic uncertainty of 0.04\,$M_\odot$ for best-fit models from individual frequency fitting, we found an agreement within 1.2$\sigma$. From the scaling relations, we obtained a mass range of 0.93 to 0.98$M_\odot$ by using the observed large frequency separations (\dnu) in the scaling relations for the primary. By using standard corrections for departures from the asymptotic regime of \dnu, we obtained a mass range of 0.83 to 1.03$M_\odot$. The upper ends of both ranges agree well with the dynamical mass of the primary. This approach provides the first empirical validation for MS solar-like oscillators and opens a new window for validating the asteroseismology. Through a dedicated program targeting   astrometric SB2 binary systems, ESA's PLATO space mission will effectively enlarge the benchmark sample to a considerable extent.
}


\let\linenumbers\nolinenumbers\nolinenumbers

\maketitle

\linenumbers\modulolinenumbers[5]

\section{Introduction}

A star's mass is the fundamental factor governing its structural properties and evolutionary timescales \citep[][and references therein]{Serenelli2021}. However, it is challenging to determine this fundamental parameter with precision and accuracy. Asteroseismology is a powerful technique for determining stellar mass by analyzing the pattern of oscillation modes in the power spectral density (PSD).
The exquisite photometric quality of space telescopes such as the 
 Convection, Rotation \& planetary Transits  \citep[CoRoT,][]{Baglin2009}, 
NASA's \Kepler\ and K2 \citep[resp.]{Borucki2010,Howell2014}, and Transiting Exoplanet Survey Satellite \citep[TESS,][]{Ricker2014} missions have provided valuable data on hundreds of thousands of oscillating stars \citep[][]{Garcia2019LRSP}.

Stellar masses can be determined by exploiting the power excess of convectively excited solar-like oscillating stars and  there are two  approaches that are most commonly used for this purpose. If the PSD contains a sufficient number of high-quality oscillation modes with identified spherical degrees ($\ell$), we can search for the best-fitting stellar model by modeling the individual frequencies \citep{LebretonGoupil2014, 
Mathur2012,
Metcalfe2014,
Serenelli2017, 
Creevey2017, 
LiTanda2024,
Buldgen2025}. 
The stellar radius ($R$), mass ($M$), and age inferred from the best-fitting model are typically reported with uncertainties of 
$\sim$1\%, $\sim$4\%, and $\sim$11\%, respectively \citep[e.g.,][]{Cunha2021}. 
However, detailed individual frequency modeling requires high-fidelity mode characterization, requiring significant computational resources, which increase in line with the star's progressive stage of evolution. Therefore, the second, global approach for seismic parameter determination becomes relevant: asteroseismic scaling relations.

These scaling relations, formulated by \cite{Brown1991}, \cite{Kjeldsen1995}, and \cite{kallinger2010}, 
\begin{align}
\frac{R}{R_\odot} &\simeq \left(\frac{\nu_{\rm max}}{\nu_{\rm max}^\odot}\right) \left(\frac{\Delta\nu}{f_{\Delta\nu}\,\Delta\nu_\odot}\right)^{-2} \left(\frac{T_{\rm eff}}{T_{\rm eff}^\odot}\right)^{1/2} \label{eq:radius} ,\\
\frac{M}{M_\odot} &\simeq \left(\frac{\nu_{\rm max}}{\nu_{\rm max}^\odot}\right)^{3} \left(\frac{\Delta\nu}{f_{\Delta\nu}\,\Delta\nu_\odot}\right)^{-4} \left(\frac{T_{\rm eff}}{T_{\rm eff}^\odot}\right)^{3/2} \simeq \left(\frac{R}{R_\odot}\right)^3 \left(\frac{\Delta\nu}{f_{\Delta\nu}\,\Delta\nu_\odot}\right)^2, \label{eq:mass}
\end{align}
provide estimates of the mass and radius of the solar-like oscillators in solar units \citep[see also][]{Stello2009a, Sharma2016, Hekker2020, Belkacem2013}. 
These equations use the frequency of oscillatory power excess ($\nu_{\rm max}$), large-frequency separation ($\Delta\nu$) between modes of identical $\ell$, consecutive radial orders ($n$), and effective temperature ($T_\mathrm{eff}$) measured from the star and the Sun ($\odot$) as input for these homological relations. The factor, $f_{\Delta\nu}$, corrects for deviations of the observed value of \dnu.
These relations are widely applied on large samples in galactic archaeology, stellar evolution, and exoplanet host characterization \citep[e.g.,][]{Casagrande2016, 
Mathur2016, 
Pinsonneault2025, 
Huber2022}. 
Typically a precision of $\sim$3\% in radius and $\sim$6\% in mass  are reported from large sample studies \citep[e.g.,][]{Serenelli2017}. 
However, both methods still lack robust comparison with model-independent mass estimates, particularly on the MS.

The only direct test for seismic masses on the MS was performed by \cite{Gaulme2016Sun}, by studying the actual oscillation of the Sun, using its reflection on Neptune from K2 photometry. The authors found the solar mass and radius to be overestimated by $\sim$14\% and $\sim$4\%, respectively, 
from the scaling relations. They attributed this discrepancy to shifts in frequency due to enhanced solar activity. Indeed, the best-fit 
model constrained through individual frequencies led to a mass and radius consistent with simultaneous observations from the ESA Solar and Heliospheric Observatory \citep[SOHO,][]{Domingo1995}. 
The lack of benchmark systems has left a critical gap in validating seismic masses precisely where they are most often assumed to be robust. 

Eclipsing SB2 systems provide model-independent measurements of the stellar masses, constrained from the orbital motion \citep[e.g.,][]{Prsa2018}. Comparisons of the dynamical mass with the seismically inferred mass for red giants have suggested that asteroseismology could end up overestimating the stellar mass by up to 15\% \citep{Gaulme2016}. Such a dichotomy between the seismic and dynamical masses has been further discussed by several other papers \citep[e.g.,][]{ Benbakoura2021, Li2022}. Despite large space-photometry datasets, the benchmark sample of eclipsing SB2 systems with solar-like oscillators remains small ($\sim$20) and is heavily biased toward H-shell burning red giants \citep{Beck2024}. While $\sim$30\% of all known (MS) solar-like oscillators are in binary systems \citep{Beck2025}, only few binaries have been  seismically studied \citep[][]{Kjeldsen2005, White2017, Metcalfe2012, Beck2017Lithium}

The key challenge in the MS mass regime (0.8\,$\lesssim$\,$M/M_\odot$\,$\lesssim$1.6, K5-F5) has been the lack of systems with known orbital inclinations, which are necessary to derive dynamical masses. Eclipses naturally provide this constraint from their light curve models. However, the third data release \citep[DR3;][]{GaiaDR3Vallenari2022} from ESA's \textit{Gaia} mission \citep{ESAGaiaPrusti2016} has opened a new avenue. For a subset of systems, the \Gaia non-single star catalog \citep[NSS,][]{Arenou2023} provides orbital inclinations for astrometric and single-lined binaries (ASB1). 
While the inclination is normally the bottleneck for the determination of dynamical masses, for the case of astrometric systems, this parameter is known. Thus, the missing piece becomes the detection of the spectroscopic signature of the secondary component from ground-based observations.

In this Letter, we report on the G1V star  KIC\,9693187\footnote{=\,Gaia\,DR3\,2107491287660520832\,=\,2MASS\,J18510009+4625209}, the least evolved solar-like oscillator so far,
whose dynamical mass can be derived through a combination of \textit{Gaia}\,DR3 astrometry and ground-based SB2 spectroscopy from the Las Cumbres Observatory Global Telescope Network \citep[LCO,][]{Brown2013LCO}. 
\cite{Chaplin2014} identified and characterized as a solar-like oscillator 
and the color-magnitude diagram (CMD) suggests a position 
on the MS or an early subgiant (eSG) \citep{Godoy-Rivera2025}.
\cite{MolendaZakowicz2013} reported this system as an SB2 system, but did not provide orbital parameters. Eventually,  \cite{Arenou2023} reported this system as ASB1 within the framework of Two-Body Orbit solutions (TBO) of the NSS. In addition to the orbital parameters (\hbox{$P_\mathrm{orb}$=103.75$\pm$0.13\,days}, \hbox{$e$=0.39$\pm$0.04}), the TBO also reports the inclination angle, $i$, of the orbital plane to be 66$\pm$2\,degrees.

\vspace{-3mm}
\section{Spectroscopic analysis and dynamical masses \label{sec:Spectroscopy}}

We monitored \KIC with Network of Robotic Echelle Spectrographs 
\citep[NRES,][]{Brown2013LCO}, mounted on the 1\,m telescopes of the northern LCO nodes at McDonald Observatory (USA) and Wise Observatory (Israel). NRES is an échelle spectrograph with a resolution of $R$\,$\simeq$\,53,000. 
Between April 2023 and June 2024, we have collected 22 spectra of \KIC.

The NRES pipeline calculates the cross-correlation function's response profile (CCF) from the observed spectra and a synthetic spectrum as a template to compute the radial velocities (RV). The pipeline automatically selected the most appropriate template from the Phoenix model library, based on the \Gaia parameters of the target ($T_\mathrm{eff}$\,=\,5700\,K, $\log g$\,=\,4\,dex, [Fe/H]\,=\,0\,dex). The CCF profiles of the individual spectra, shown in Fig.\,\ref{fig:RV2} clearly reveal the SB2 nature of \KIC.

\vspace{-2mm}
\subsection{Spectral disentangling \label{sec:disentangling}}

To determine the RV semi-amplitudes of the primary and secondary, $K_1$ and $K_2$, respectively, we applied spectral disentangling (SPD) in Fourier space.  
This technique optimizes the orbital elements of a binary, including $K_1$ and $K_2$  \citep{SimonSturm1994,Hadrava1995}. We note that we refer to the more massive component as the primary, indicated with the subscript 1.

For our SPD analysis of \KIC, we use the {\sc FDBinary} code \citep{Ilijic2004}. 
We focused on a 60\,nm wide region between 476.1 and 536.1\,nm, including the Mg-triplet, which is well suited to determining $K$ for both components. The disentangled spectra are depicted in Fig.\,\ref{fig:DisentangledSpectra}. 
Except for the orbital period, which we fix to the astrometric solution from the \Gaia TBO, we treated all orbital parameters as free parameters. The resulting values are reported in Table\,\ref{tab:DisentanglingSolution}. The uncertainties were derived from a Gaussian fit to the parameter distributions from 5000 bootstrap simulations \citep[see][]{Pavlovski2023}. 

The best solution finds $K_1$=26.298\,$\pm$0.058\,km/s
and $K_2$=29.394\,$\pm$0.066\,km/s
as the RVs semi-amplitudes for the primary and secondary component.  These values reveal a mass ratio of  $q$\,=\,$M_2/M_1$\,=0.8947\,$\pm$0.0028 ($\sim$11\% difference in mass).

\vspace{-2mm}
\subsection{Stellar fundamental parameters} 
Fundamental stellar parameters were derived through a spectroscopic analysis using an updated version of the 1D/LTE code \texttt{BACCHUS} \citep{2016ascl.soft05004M, Hayes2022}. The full analysis and derived parameters are provided in Appendix\,\ref{sec:AppendixA_fundParameters} and Table\,\ref{tab:DisentanglingSolution}. 
We find an effective temperature of 5738\,$\pm$84K and a subsolar metallicity ([Fe/H]=-0.36\,$\pm$0.15dex). The significant enrichment of  $\alpha$ elements ([$\alpha$/Fe]=+0.25\,$\pm$0.05dex) indicates that the star is member of the older thick-disk population.
The projected surface rotation velocity, $v\sin i$, of the primary and secondary (6.7$\pm$1.5, and 8.0$\pm$2.0 km/s, respectively) are more than three times the current solar value of $v\sin i_\odot$\,$\simeq$\,1.96\,km/s
\citep{Beck2005} and about six times faster than the expected $v\sin i$ at the MS turnoff of the Sun.
Even more puzzling, no photometric signature of spot-modulation is seen in the light curve or PSD \citep[][see Fig.\,\ref{fig:globalSeismology} below 1\,$\mu$Hz]{Garcia2014,Santos2021}.

\begin{table}[t]
\caption{Orbital and fundamental parameters for \KIC. \vspace{-2mm}}

\tabcolsep=9pt
\centering
\begin{tabular}{cc|rr}
\hline\hline
\multicolumn{2}{c}{Parameter} & Primary & Secondary \\ \hline
$P_\mathrm{orb}$ &  [d]           & \multicolumn{2}{c}{103.75\,$\pm$0.13 (fixed)}\\
$e$ &             & \multicolumn{2}{c}{0.3926\,$\pm$0.0016}\\
$\omega$ & [deg]     & \multicolumn{2}{c}{225.98\,$\pm$0.53}\\
$K$ & [km/s]      & 26.298\,$\pm$0.058 &29.394\,$\pm$0.066\\
$q$ & =\,$M_2/M_1$    & \multicolumn{2}{c}{0.8947\,$\pm$0.0028}\\
$T_0$ & MJD        & \multicolumn{2}{c}{60020.40\,$\pm$0.11 }\\
\hline
$M_1^\mathrm{dyn}$ & [M$_\odot$] &
0.993\,$\pm$0.046 &
 0.889\,$\pm$0.041 \\

\hline
\end{tabular}\vspace{-7mm}
\tablefoot{The table notes are provided in Appendix\,\ref{tablenotes:disentangling}.\vspace{-15mm}
}
\label{tab:DisentanglingSolution}

\end{table}

\vspace{-1mm}
\subsection{Dynamical masses \label{sec:DynamicalMasses}}
The dynamical masses for both components are given by
\begin{equation}
\frac{M_{1,2}}{M_\odot} = \frac{K_{2,1} \cdot (K_1 + K_2)^2}{M_\odot  G} \cdot \frac{P \cdot (1 - e^2)^{3/2}}{2 \pi \cdot \sin^3 i}\,.
\label{eq:dynamicalMasses}
\end{equation}

From the determined orbital parameters and $K_{1,2}$ values taken from spectroscopy (Table\,\ref{tab:DisentanglingSolution}),  
along with the $P_\mathrm{orb}$ and inclination, $i$, provided by \textit{Gaia}, we obtained 
$M^\mathrm{dyn}_1$=0.993$\pm$0.046\,M$_\odot$ and 
$M^\mathrm{dyn}_2$=0.889$\pm$0.041\,M$_\odot$.

\section{Asteroseismology of \KIC \label{sec:Seismology}}

\subsection{Global seismology}

To test the asteroseismic scaling relations for the mass (Eq.\,\ref{eq:mass}), we used the effective temperature and the global seismic parameters of the oscillating component, obtained with the A2Z pipeline and reported by \cite{Mathur2022}. Furthermore, we re-extracted the light-curve and calculated the PSD (for details see Appendix\,\ref{sec:stitching} and Fig.\,\ref{fig:PSD}) and redetermined the values using the universal pattern module of the \texttt{apollinaire} peakbagging code \citep[][]{Breton2021}, as described in Appendix\,\ref{sec:ScalingRelations}. The values from both methods agree within the uncertainty of the respective parameters (see Table\,\ref{tab:globalSeismology}). However, \texttt{apollinaire} leads to smaller uncertainties and a lower value for \num. Figure\,\ref{fig:scalingRelationComparison} compares the results of the scaling relations,  without performing a correction on \dnu, and following the correction by \cite{Mosser2013} and  \cite{Li2023} (see Appendix\,\ref{sec:ScalingRelations}) with the primary's dynamical mass.

\subsection{Individual-frequency seismology}
Using \texttt{apollinaire},
we extracted 28 frequencies 
of the spherical degree, $\ell$\,=\,0,\,1,\,and\,2.
The extracted frequencies (see Table\,\ref{tab:individualFrequencies}) are depicted in the échelle diagram~(Fig.\,\ref{fig:globalSeismology}).
We searched for the best stellar model, constrained by using individual-frequency (IF)  seismology.  
The structural modeling was done with the Modules for Experiments in Stellar Astrophysics stellar evolution code package \citep[MESA,][and references therein]{MESA2023,Paxton2011}. The frequencies were calculated from each model using the Aarhus adiabatic
pulsation codes \citep{JcD2008}. 
The modeling setup and input parameters are described in  detail in Appendix\,\ref{sec:AppB_mesa}.
The model frequencies in the échelle diagram (Fig.\,\ref{fig:globalSeismology}) present a good fit.
The model that fits best the observational constraints is found with 
$M^\mathrm{IF}_1$=0.92$\pm$0.01\,M$_\odot$ and  
$R^\mathrm{IF}_1$=1.38$\pm$0.01\,R$_\odot$ (see Table\,\ref{tab:globalSeismology}). 

\section{Discussion and conclusions
\label{sec:DiscussionsConclusions}}

\begin{figure}
    \centering
    \includegraphics[width=\columnwidth,height=45mm]{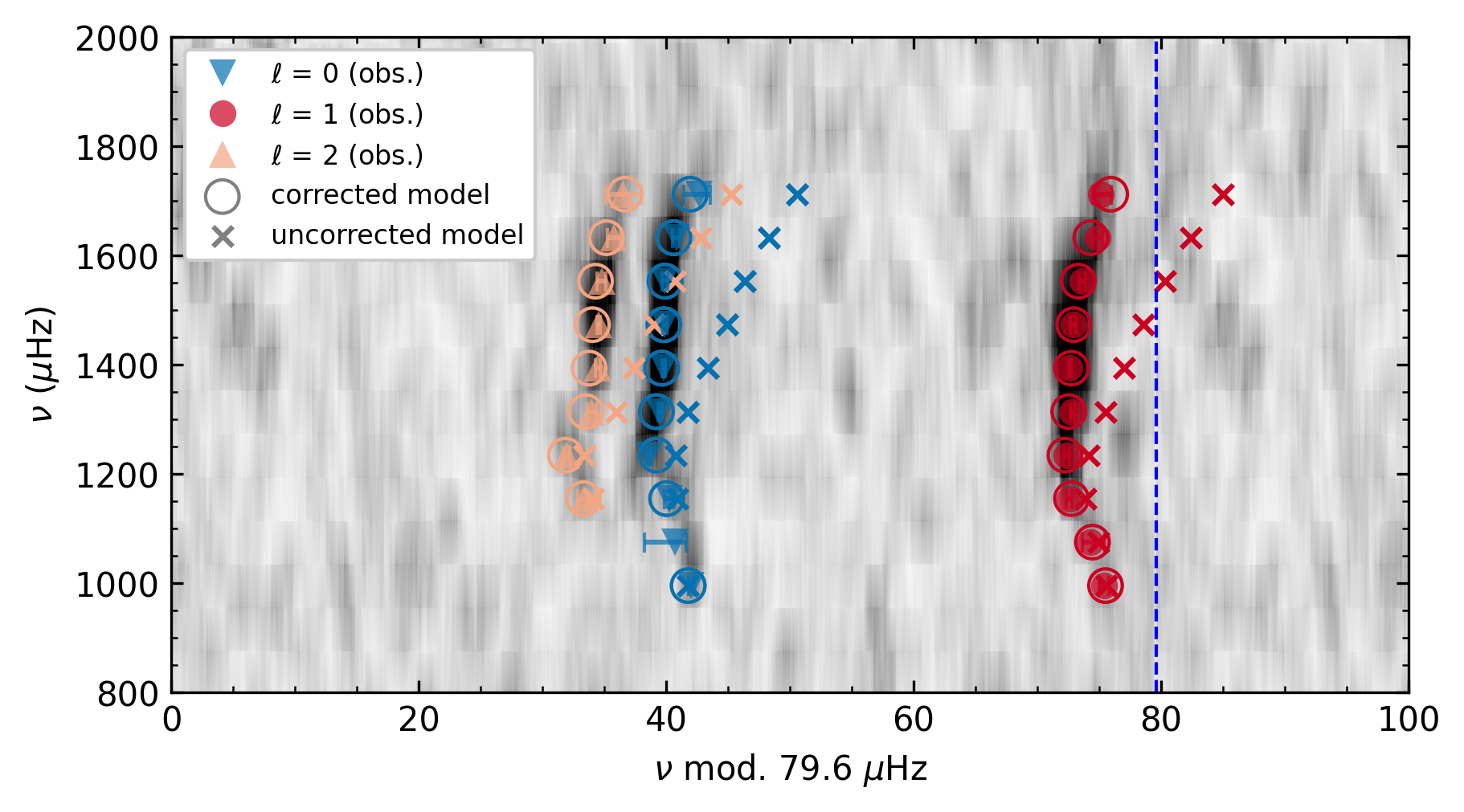}
    \vspace{-7mm}
    \caption{Echelle diagram of the target for the  observed (closed symbols) and modeled (open symbols) oscillation modes. Crosses mark the positions of the model frequencies, uncorrected for surface effects.}
    \label{fig:globalSeismology}
    
    \vspace{3mm}
    \includegraphics[width=\columnwidth,height=45mm]{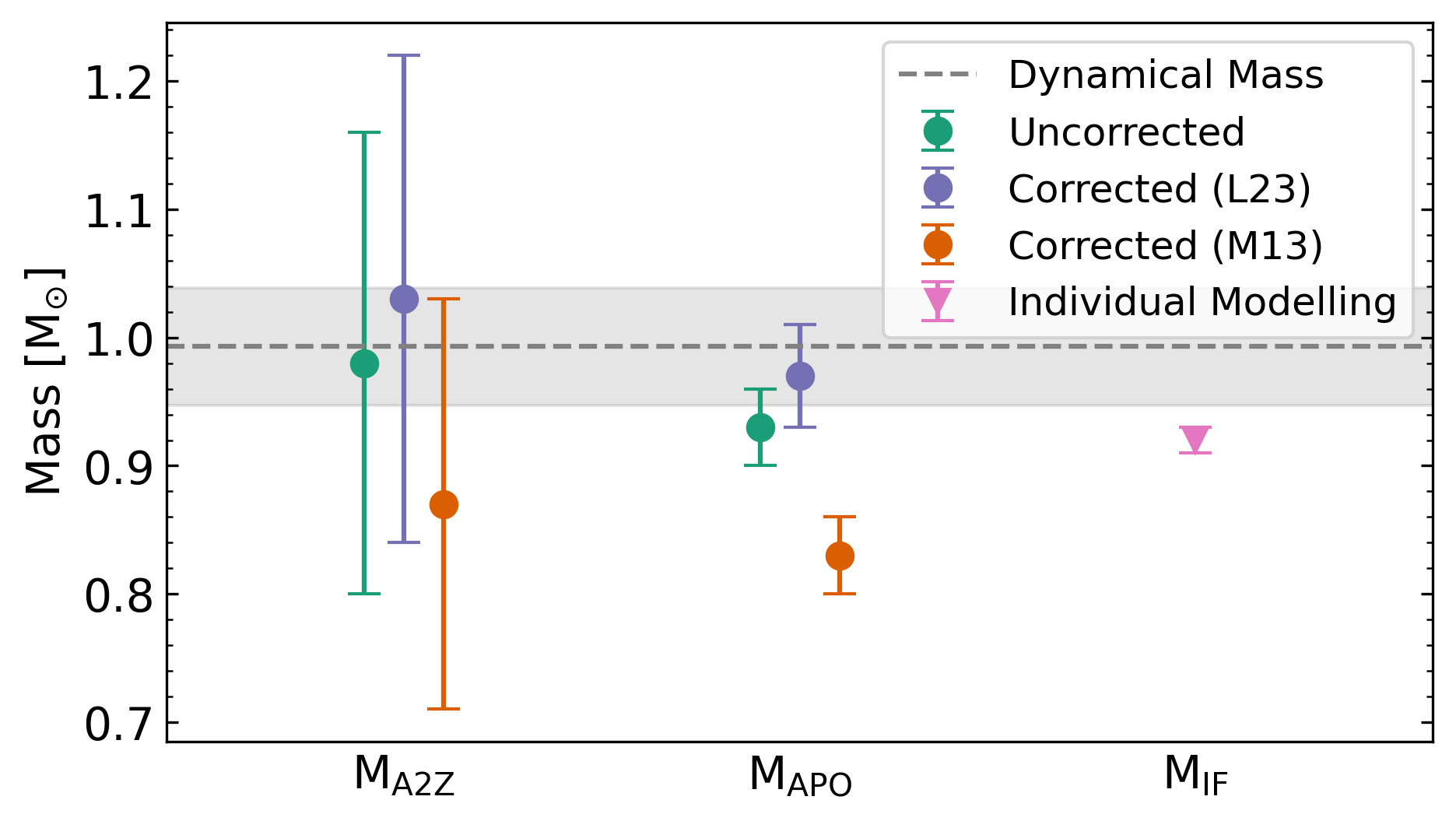}
    \vspace{-3mm}
    \caption{Mass estimates of the primary of \KIC from asteroseismic scaling relations, corrected following \citet[purple]{Li2023}, \citet[orange]{Mosser2013}, and uncorrected (green), as well as the individual frequency modeling (rose). The horizontal line and grey-shaded area represent the dynamical mass its related uncertainties, respectively.}
    \label{fig:scalingRelationComparison}

\end{figure}

Figure\,\ref{fig:scalingRelationComparison} and compares the primary's dynamical mass with seismically inferred masses from scaling relations and individual frequency modeling (Table\,\ref{tab:globalSeismology}). From scaling relations, we obtained a mass range of 0.93$\pm$0.03$M_\odot$ to 0.98$\pm$0.18$M_\odot$ by using the observed \dnu from \apol and A2Z in the scaling relations for the primary. While the masses from the A2Z values agree within $\sim$2\%, they suffer from large uncertainties. The \apol values lead to a mass that is $\sim$6\% below, but still within the uncertainties of the dynamical mass.

The large-frequency separation is known to depart from the asymptotic regime increasingly with a decreasing radial order. To correct the obtained masses, we tested two well-established formalisms from the literature (for details see Appendix\,\ref{sec:ScalingRelations}).
The corrections proposed by \cite{Mosser2013}
appear to underestimate the seismic mass for this late MS target by $\sim$23\%, based on the \apol set of parameters.
The best agreement ($\sim$2\%) is found from the correction of \cite{Li2022}.

The mass obtained from detailed seismic modeling using individual frequencies, \( M^\mathrm{IF}_1 = 0.92 \pm 0.01\,\mathrm{M}_\odot \), agrees with the dynamical mass within 1.55\(\sigma\), where \(\sigma\) denotes the combined uncertainty from both measurements. The best-fitting seismic model reproduces the observables well, as indicated by a reduced \(\chi^2_\mathrm{red} = 1.53\) and the corresponding échelle diagram shown in Fig.\,\ref{fig:globalSeismology}.
The small formal uncertainty in the IF seismic mass reflects internal modeling precision. However, results from the hare-and-hounds exercise by \citet{Cunha2021} suggest typical systematic deviations of up to 4.32\% in mass, 1.33\% in radius, and 11.25\% in age. Accounting for such systematics, the uncertainty in the IF seismic mass would increase to approximately \(0.04\,\mathrm{M}_\odot\), leading to a difference of 1.2\(\sigma\) relative to the dynamical value. 
Therefore, the level of agreement between the dynamical and seismic masses remains robust.

The best-fit IF model yields an age of 11.20\,$\pm$0.55\,Gyr for the primary component of \KIC\ (Table\,\ref{tab:globalSeismology}). This result confirms the evolved nature of the star, placing it at the main sequence (MS) turnoff, at the point of core hydrogen exhaustion, while the secondary remains on the MS (Fig.\,\ref{fig:HRD}). This evolutionary configuration is consistent with the system's $\alpha$-enhanced chemical composition. However, given the primary's advanced age, the rapid surface rotation remains puzzling.

The uncertainty on the derived dynamical masses remains relatively large compared to the subpercent precision that is in principle achievable with this method \citep{Torres2010AccurateMasses}. A major contributor is the relatively large error of \(\pm 2^\circ\) on the orbital inclination, which significantly contributes to the 
uncertainty. Looking ahead, larger sample sizes with improved constraints on orbital inclination and the availability of epoch RV in the forthcoming \textit{Gaia}\,DR4, potentially combined with ground-based RV monitoring, will enable rapid progress in refining dynamical mass measurements for solar-like oscillators in binaries. 

As suggested by \citet{Beck2024}, the combination of \textit{Gaia} astrometry of non-eclipsing systems with SB2 solutions from ground-based spectroscopy (or from forthcoming \Gaia DRs) offers a promising and abundant new source of benchmark calibrators. These developments are particularly timely in light of the upcoming ESA mission PLAnetary Transits and Oscillations of stars \citep[PLATO,][]{Plato2025}, which will provide high-precision asteroseismic data for ten thousands of solar-like oscillators. A significant number of ASB1 systems identified in \GaiaDR as astrometric binaries, potentially hosting a solar-like oscillator on the MS, SG,  or red-giant phase, have been selected by the authors as part of the Science Calibration and Validation PLATO Input Catalog (scvPIC).
The NRES/LCO monitoring presented in this letter is part of a dedicated follow-up program to build an SB2 sample of such high-value targets. This growing number of well-calibrated benchmark stars enables a broader validation of global asteroseismic methods, offering a rich ensemble of well-calibrated stars to test the intricacies of stellar evolution from physics-informed models \citep[e.g.,][]{Grossmann2025,Thomsen2025, Schimak2026}. Ultimately, these efforts will support PLATO’s core scientific objective of improving the precision of stellar age determinations.

\begin{acknowledgements}
The authors thank the people behind the ESA \Gaia, NASA \Kepler mission and the LCOT project and acknowledge the contribution of the IAC High-Performance Computing support team and hardware facilities to the results of this research. The authors thank Dr.\,Dennis\,Stello for comments in the TASC community review that improved the paper. 
PGB acknowledges support by the Spanish Ministry of Science,\,Innovation\,\&\,Universities (MCIN) with the \textit{Ram{\'o}n\,y\,Cajal} fellowship (RYC-2021-033137-I, MRR4032204). PGB,\,TM,\,DGR,\,DHM,\,JM,\,\&\,IA acknowledge support from the MCIN project \textit{PLAtoSOnG} (PID2023-146453NB-100, PI:\,Beck) and DHG,\,SM,\,DGR,\,RAG\,\&\,HJD acknowledge support from the MCIN grant PID2023-149439NB-C41. 
DGR acknowledges support from the \textit{Juan\,de\,la\,Cierva} program under contract JDC2022-049054-I. 
DHG received support from the \textit{”la\,Caixa”\,Foundation} fellowship (ID\,100010434, LCF/BQ/DI23/11990068).  JM was supported by the Czech Science Foundation (GAČR), project \#24-10608O. DBP,\,RAG,\,\&\,EP acknowledge the support from the GOLF\,\&\,PLATO Centre National D'{\'{E}}tudes Spatiales grants. 
YL acknowledges support from STFC studentship ST/W507453/1 and ERC Consolidator Grant GAIA-BIFROST (Grant\,ID\,101003096).
This work was supported by the {International Space Science Institute (ISSI) in Bern (project\,24-629).}\\

This article is based on observations made with the NRES spectrograph mounted on the 1\,m telescopes LCO telescopes, one of whose nodes is located at the Observatorios de Canarias del IAC at Teide Observatory.
This work has made use of data from the ESA mission
\href{https://www.cosmos.esa.int/gaia}{\Gaia}, processed by the \href{https://www.cosmos.esa.int/web/gaia/dpac/consortium}{\Gaia
Data Processing and Analysis Consortium (DPAC)}. Funding for the DPAC
has been provided by national institutions, in particular the institutions
participating in the \Gaia Multilateral Agreement. 
Data collected with the \textit{Kepler} mission, obtained from the MAST data archive at the STScI was used. 
STScI is operated by the Association of Universities for Research in Astronomy, Inc.(NASA contract NAS\,5–26555).

\end{acknowledgements}
\vspace{-7mm}

\bibliographystyle{aa.bst}
\bibliography{aa57452-25.bib}

\begin{appendix}
\section{Spectroscopic analysis }

\subsection{Fundamental parameters 
\label{sec:AppendixA_fundParameters}}\vspace{-2mm}

Fundamental stellar parameters were derived through a spectroscopic analysis using an updated version of the 1D/LTE code \texttt{BACCHUS} \citep[Brussels Automatic Code for Characterizing High accUracy Spectra,][]{2016ascl.soft05004M, Hayes2022}, which employs MARCS model atmospheres \citep{Gustafsson2008} and the atomic and molecular line lists from \citet{Heiter2021}. For the specific case of SB2 spectra, the code has been newly extended to compute a secondary synthetic spectrum on the fly, with fixed stellar parameters, a given radius ratio, and the appropriate radial velocity shift. This secondary spectrum is treated as a perturbation to the primary spectrum, adding continuum opacities of both components, and possibly line opacities when analyzing non-disentangled spectra. Thus, from the perspective of the code, such implementation allows to indifferently swap the analysis of each component. 

Effective temperatures were determined by requiring no trend between the abundances derived from \ion{Fe}{I} lines and their excitation potentials. Surface gravity was obtained by enforcing ionization balance between \ion{Fe}{I} and \ion{Fe}{II} lines. As this method requires a large number of moderately weak Fe lines, we obtained and combined three consecutive spectra with high signal-to-noise.
Microturbulence velocity was simultaneously constrained by minimizing trends between Fe line abundances and their equivalent widths. The metallicity was determined as the mean abundance of the \ion{Fe}{I} lines. However, because the secondary component of KIC~9693187 contributes significantly less to the overall flux, only the strongest lines could be observed and analyzed. Consequently, neither the microturbulence nor the surface gravity could be constrained with our procedure relying on Fe lines and they were fixed in the determination of the parameters.  The entire parameter determination process was iteratively refined, alternating the primary and secondary components analysis until convergence was reached. The temperatures of the two components were also validated by checking the quality of the fit of the wing of H$\alpha$ line of the disentangled spectra.

Moreover, in SB2 systems, on a first order approximation the line-depth-ratio, thus metallicity ratio, and luminosity ratio are degenerate. We stress though, that the clear orbital motion of both components supports the assumption of a common origin and homogeneous chemical composition, as expected for coeval binaries \citep[e.g.,][]{GodoyRivera2018, MoeStefano2017, Offner2022}. Therefore, by enforcing equal metallicities, luminosity could be constrained. On another hand, absorption line depth may also dependent on other stellar parameters such as the effective temperature. 
To alleviate the stellar parameters dependence, very strong and saturated lines such as the core of the Mg triplet lines represent relevant features which depth is nearly invariant with temperature, surface gravity or metallicity in cool stars spectra. 
We stress that the core of Mg lines are also subject to NLTE effects \citep{Osorio2016,Alexeeva2018}. Caution must be taken when {using those lines as luminosity indicator.} 

\begin{figure}[t!]
    \centering
    \includegraphics[width=\columnwidth, height=45mm]{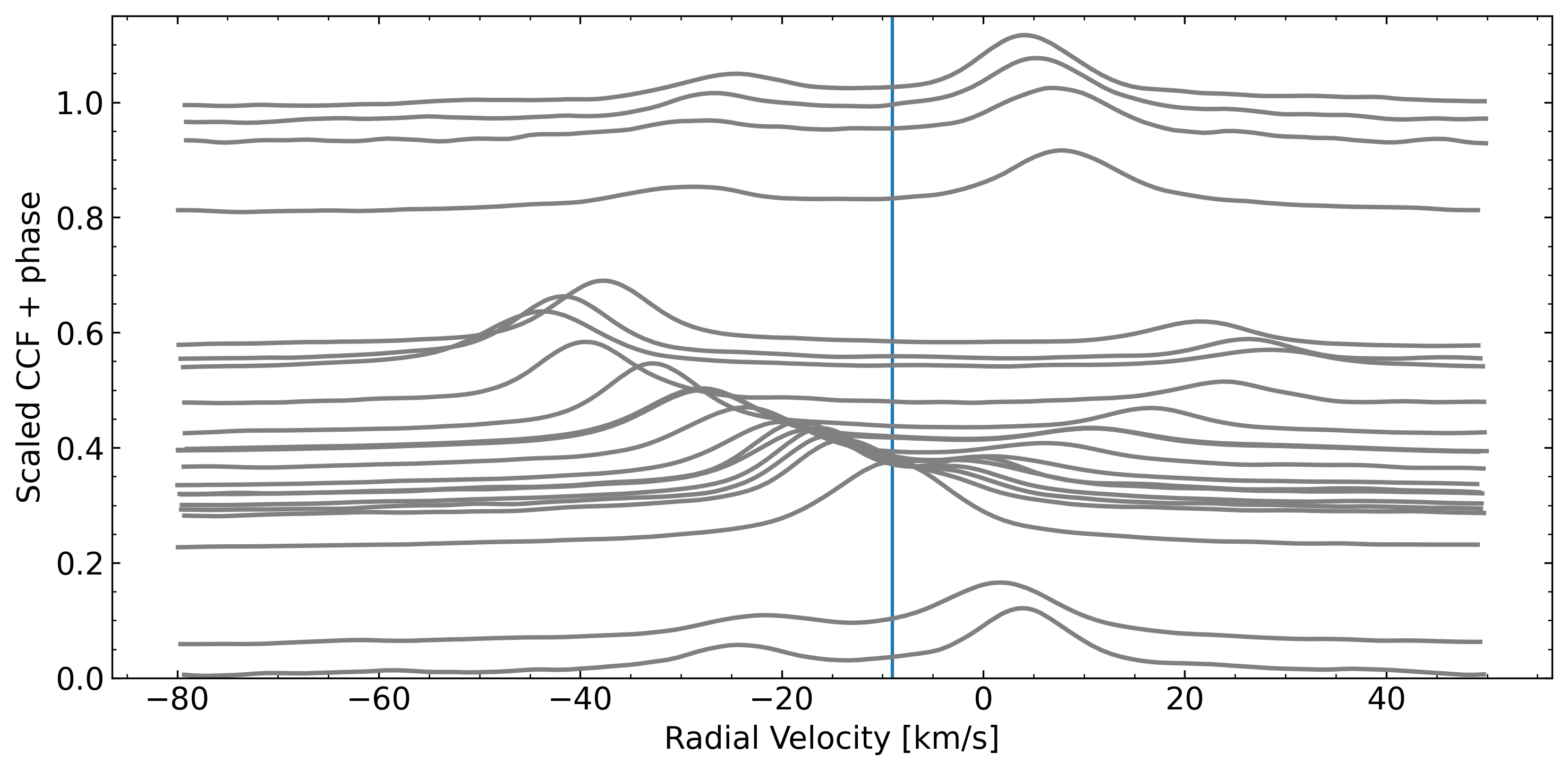}
    \caption{CCF profiles for \KIC as function of the orbital phase. The vertical line marks the systematic velocity of the system. 
    }
    \label{fig:RV2}
\vspace{3mm}
    \centering 
    \vspace{-1mm} 
    \includegraphics[width=\columnwidth,height=46mm]{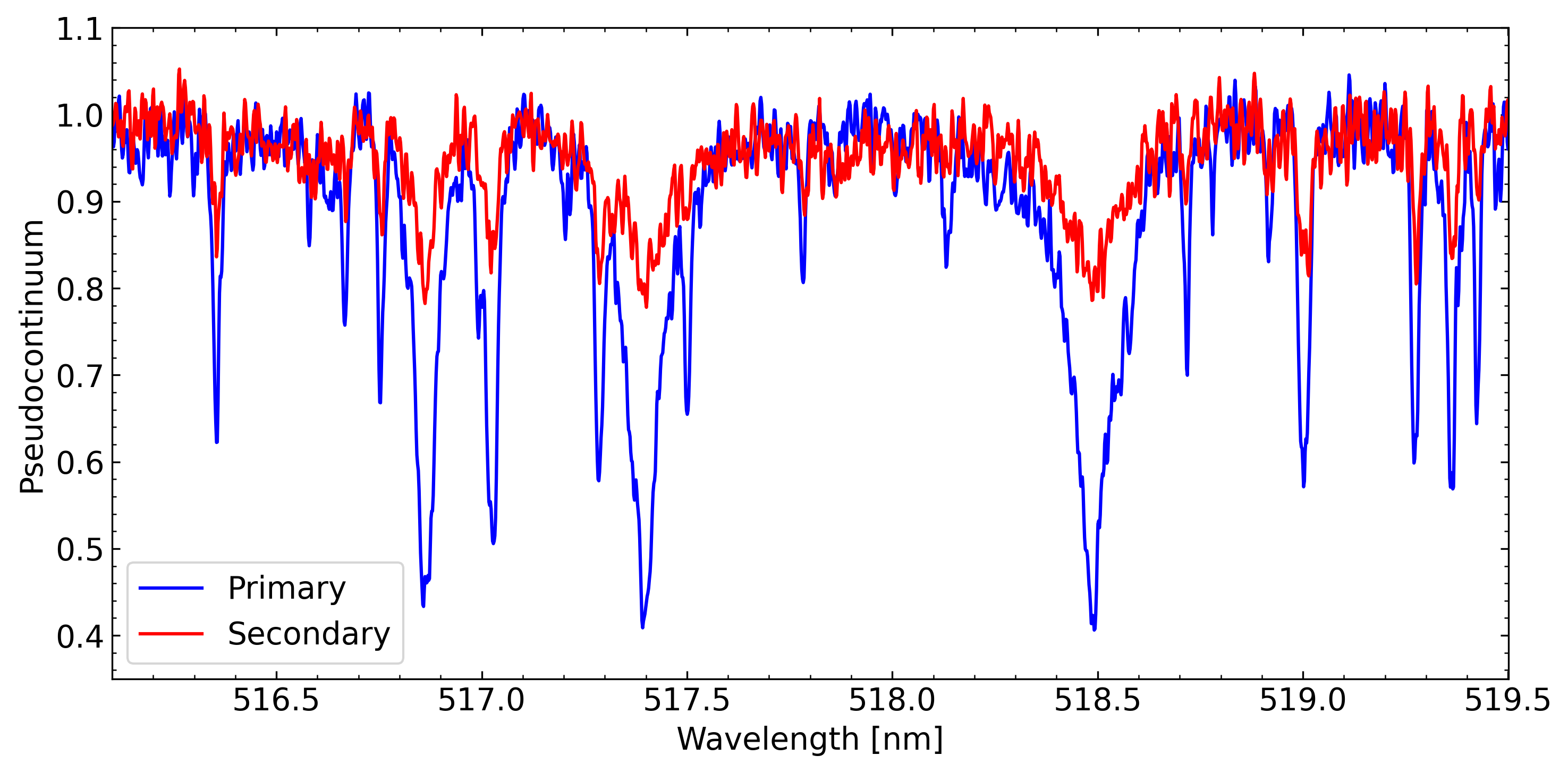}
    \caption{Accumulated spectra for the primary (blue) and secondary (red) of \KIC around the Mg triplet from spectral disentangling. The disentangled  spectra are both normalized to the pseudocontinuum of the composite spectrum. }
    \label{fig:DisentangledSpectra}
\end{figure}

Finally, with both temperature of the two component constrained and the luminosity ratio fixed, the radii ratio could be established \citep{Masseron2012}, leading to a value of 1.4$\pm$0.05.
From the line-depth ratio in the disentangled wavelength  range (Fig.\,\ref{fig:DisentangledSpectra}) of about $\sim$68:$\sim$32, which is a proxy for the fractional-light ratio between both components in Johnson\,V.

The code also automatically adjusts a mean line-broadening value to fit the Fe line profiles. The projected rotational velocity ($v\sin i$) was then estimated by assuming that this broadening is the quadratic sum of the instrumental resolution ($R = 53{,}000$), macroturbulence (adopted from the prescription of \citet{Doyle2014} for G–K MS stars), and rotational broadening. The derived fundamental parameters are reported in Table\,\ref{tab:DisentanglingSolution}.

\begin{table}[t]
\caption{Orbital and fundamental parameters for \KIC. \vspace{-2mm}}

\tabcolsep=9pt
\centering
\begin{tabular}{cc|rr}
\hline\hline
\multicolumn{2}{c}{Parameter} & Primary & Secondary \\ \hline
$T_\mathrm{eff}$ & [K] & 5738\,$\pm$84& 5150\,$\pm$150 \\
$\log g$ & [dex] & 4.00\,$\pm$0.37 & 4.5 (fixed)\\
$v\sin$i& [km/s] & 6.7\,$\pm$1.5 & 8\,$\pm$2.0 \\
$\xi_\mathrm{micro}$ & [km/s] & 1.21\,$\pm$0.07 & 1.0 (fixed)\\
$[$Fe/H$]$ & [dex]  & \multicolumn{2}{c}{-0.36\,$\pm$0.15}\\ 
$[\alpha$/Fe] & [dex] & \multicolumn{2}{c}{+0.25\,$\pm$0.05}\\
$[$M/H] & [dex] & \multicolumn{2}{c}{+0.18\,$\pm$0.15}\\
$L_1 / L_2$  &     & \multicolumn{2}{c}{2.6\,$\pm$0.1} \\
\hline
\end{tabular}\vspace{-2mm}
\tablefoot{The table notes are provided in Appendix\,\ref{tablenotes:spectroscopy}.
\vspace{-12mm}
}
\label{tab:spectroscopicParameters}
\end{table}

\vspace{-2mm}
\subsection{Notes on Table\,\ref{tab:DisentanglingSolution}\label{tablenotes:disentangling}}
\vspace{-1mm}

Table\,\ref{tab:DisentanglingSolution} presents the results from spectral disentangling of the observed spectra. The bottom panel reports the dynamical masses derived from the orbital parameters and mass ratio. The individual parameters and their uncertainties are described as follows:
\begin{itemize}
\item T$_0$: time of periastron passage, expressed in BJD,
\item $e$: orbital eccentricity,
\item $\omega$ [deg]: argument of periastron, indicating the angle between the ascending node and the periastron point,
\item $K$ [km/s]: radial velocity semi-amplitude of each component,
\item $q$: mass ratio of the binary, defined as $q = M_2 / M_1$, where $M_1$ and $M_2$ are the masses of the primary and secondary,
\item $M_\mathrm{dyn}$ [M$_\odot$]: dynamical mass of each component, calculated following the formulation of  \cite{HilditchMasses}, and using the orbital parameters from the disentangling solution.
\end{itemize}

\subsection{Notes on Table\,\ref{tab:spectroscopicParameters}\label{tablenotes:spectroscopy}}
Table\,\ref{tab:spectroscopicParameters} presents the results from the spectroscopic analysis of the disentangled component spectra. The individual parameters and their uncertainties are described as follows:
\begin{itemize}
\item T$_\mathrm{eff}$ [K]: effective temperature of each component,
\item $v \sin i$ [km/s]: projected rotational surface velocity,
\item $[$M/H] [dex]: stellar metallicity, relative to the Sun,
\item $[\alpha$/Fe] [dex]: abundance of $\alpha$ elements,
\item $L$ [L$_\odot$]: stellar luminosity for the primary, from seismic scaling relations.
\end{itemize}

\section{Seismology and modeling \label{sec:Appendix_seismology}}
\begin{figure}[h!]
    \centering
    \includegraphics[width=\columnwidth]{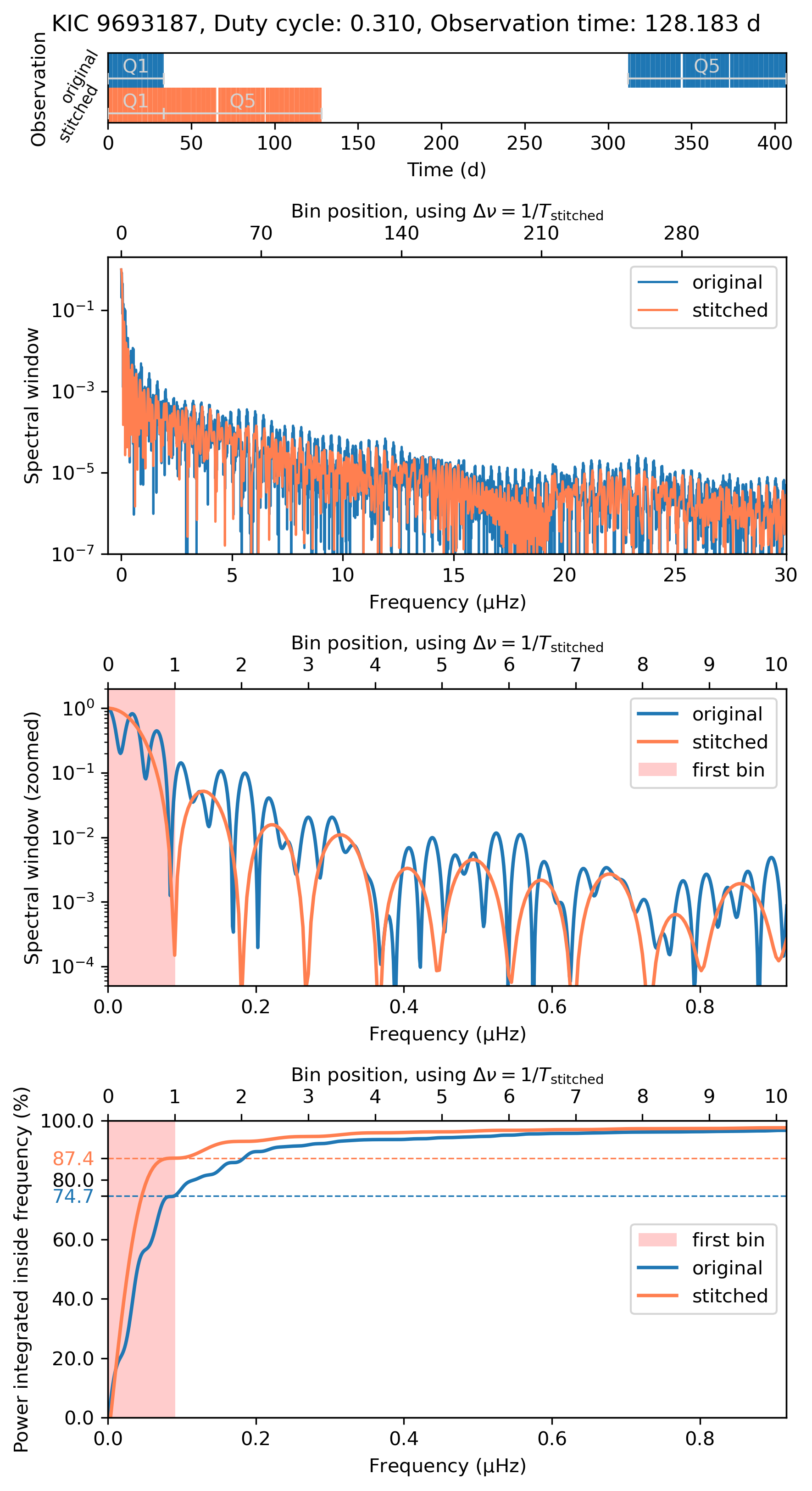}
    \caption{Comparison of the spectral window before and after removing the long gap between Q1 and Q5 for KIC~9693187. In the top panel, the window functions of the original light curve (in blue) and of the light curve after long gap removal (in orange) are shown. The second panel shows the spectral windows in both cases, and the third panel presents a zoom on the first 10 bins of the PSD, as described in the text. The bottom panel shows the cumulative power of the spectral window for both cases.}
    \label{fig:stitchedPlot}
\end{figure}

\subsection{Effect of removing the long gap between Q1 and Q5 
\label{sec:stitching}}

\KIC was observed in short cadence mode, providing one photometric measurement per minute, during two $\sim$90-day data Quarters (Q1 and Q5). 
The light curves were corrected for known systematics in the \Kepler\ photometry and the PSD was computed following the procedures of \citet[][]{garcia2011,Garcia2014} and  \citet[][]{Pires2015}.

To mitigate the impact of the window function induced by the one-year gap between observing quarters, we adopted the approach of removing this gap by shifting Q5 immediately after Q1. The impact of removing long gaps by stitching independent observing segments has been discussed by \citet{Bedding2022}, who suggested that this approach could modify the detailed shape of mode profiles and therefore affect the extracted mode frequencies and widths. To verify that the effect on the mode frequencies, which we extracted from the stitched data of \KIC, is negligible, we
computed the spectral window for both the original and the stitched light curves (Fig.~\ref{fig:stitchedPlot}, second panel). Focusing on the first ten frequency bins, defined as $1/T_{\rm stitched}$ and corresponding to the range 0–0.9\,$\mu$Hz (third panel), we find that although the main lobe of the stitched spectral window is broader by approximately a factor of three compared to the original one, the secondary lobes are significantly reduced and a larger fraction of the power is concentrated in the main lobe. This behavior is clearly quantified by the cumulative integral of the spectral window (bottom panel), which shows that the power becomes more strongly concentrated around the central frequency (87.4\% versus 74.7\%), thereby reducing spectral leakage toward higher frequencies and lowering the correlation between adjacent bins in the power spectral density.

As shown in Appendix~B of \citet{GonzalezCuesta2023}, mode frequencies extracted from a PSD computed after removing long gaps are unbiased within the quoted fitting uncertainties. In addition, reducing spectral leakage in the window function by concentrating more power into the main lobe, decreases correlations between neighboring mode peaks, which would otherwise slightly bias the uncertainties inferred by the \hbox{fitting procedure.}

\begin{table}[t]
\centering
\caption{Seismic parameters for the primary of \KIC.}
\tabcolsep=9pt
\begin{tabular}{ccrl}
\hline\hline
\multicolumn{2}{c}{Parameter} &
\multicolumn{1}{c}{Value} &
\multicolumn{1}{c}{Method/Corr.} \\[0.75mm]
\hline
$\nu_{\max}$  & [$\mu$Hz] & 1527\,$\pm$74 & A2Z\\[0.75mm]
$\Delta\nu$  &[$\mu$Hz] & 79.6\,$\pm$2.1 & A2Z\\[0.75mm]
$\nu_{\max,\odot}$  & [$\mu$Hz] & 3100 & A2Z\\[0.75mm]
$\Delta\nu_\odot$ &[$\mu$Hz] & 135.2 & A2Z\\[1.75mm]
$M^\mathrm{SR}_{1,A2Z}$ & [M$_\odot$]  &  0.98\,$\pm$0.18 & no corr. \\[0.75mm]
$M^\mathrm{SR}_{1,A2Z}$ & [M$_\odot$]  &  1.03\,$\pm$0.19 &   L23  \\[0.75mm]
$M^\mathrm{SR}_{1,A2Z}$ & [M$_\odot$]  &  0.87\,$\pm$0.16 & M13    \\[1.75mm]
$R^\mathrm{SR}_{1,A2Z}$ & [R$_\odot$]  &  1.42\,$\pm$0.10 &  no corr.\\[0.75mm]
$R^\mathrm{SR}_{1,A2Z}$ & [R$_\odot$]  &  1.45\,$\pm$0.11 & L23 \\[0.75mm]
$R^\mathrm{SR}_{1,A2Z}$ & [R$_\odot$]  &  1.34\,$\pm$0.10 &  M13\\[1.75mm]

\hline
$\nu_{\max}$ & [$\mu$Hz] & 1497\,$\pm$13 & APO \\[0.75mm]
$\Delta\nu$  &[$\mu$Hz] & 79.86\,$\pm$0.06 & APO\\[0.75mm]
$\nu_{\max,\odot}$  & [$\mu$Hz] & 3073.98\,$\pm$7.27 & APO\\[0.75mm]
$\Delta\nu_\odot$  &[$\mu$Hz] & 134.86\,$\pm$0.02 & APO\\[1.75mm]
$M^\mathrm{SR}_{1,APO}$ & [M$_\odot$] &  0.93\,$\pm$0.03& no corr.\\[0.75mm]
$M^\mathrm{SR}_{1,APO}$ & [M$_\odot$] &  0.97\,$\pm$0.04&   L23\\[0.75mm]
$M^\mathrm{SR}_{1,APO}$ & [M$_\odot$] &  0.83\,$\pm$0.30& M13 \\[1.75mm]
$R^\mathrm{SR}_{1,APO}$ & [R$_\odot$] &  1.38\,$\pm$0.02& no corr.\\[0.75mm]
$R^\mathrm{SR}_{1,APO}$ & [R$_\odot$] &  1.41\,$\pm$0.02&   L23\\[0.75mm]
$R^\mathrm{SR}_{1,APO}$ & [R$_\odot$] &  1.30\,$\pm$0.02 & M13 \\[1.75mm]

\hline
$M^\mathrm{IF}_1$ & [M$_\odot$] & {0.92\,$\pm$0.01}\\[0.75mm]
$R^\mathrm{IF}_1$ & [R$_\odot$] & {1.38\,$\pm$0.01}\\[0.75mm] 
$A^\mathrm{IF}_1$ & [Gyr] & {11.20\,$\pm$0.55}\\[0.75mm] 
\hline

\hline
\end{tabular}
\label{tab:globalSeismology}
\tablefoot{The top and middle panel report the global seismic parameters and the obtained masses and radii and their respective uncertainties for the primary for the results from the A2Z and \apol pipeline. The last column specifies the pipeline or the formalism of the \dnu correction. For details see App.\,\ref{sec:ScalingRelations}. The bottom panel reports the mass, radius and age and their respective uncertainties from the best fit model, obtained from individual frequency modeling. For details see App.\,\ref{sec:peakbagging}. 
}
\end{table}

\begin{table}
\centering
\caption{Frequency list for \KIC.}
\begin{tabular}{ccccc}
\hline\hline
n       &       $\ell$  &       $\nu$ & $\sigma_\nu$            & Flag  \\
        &               &       [$\mu$Hz]               &       [$\mu$Hz]       \\ \hline
12&0&1076.25&0.25&0\\
12&1&1109.60&0.39&0\\
13&0&1154.48&2.45&1\\
13&1&1188.09&1.15&0\\
13&2&1226.68&0.90&0\\
14&0&1233.54&0.44&0\\
14&1&1265.91&0.27&0\\
14&2&1304.83&0.29&0\\
15&0&1311.49&0.17&0\\
15&1&1345.21&0.23&0\\
15&2&1386.39&0.36&0\\
16&0&1391.91&0.18&0\\
16&1&1425.23&0.16&0\\
16&2&1466.42&0.21&0\\
17&0&1471.75&0.14&0\\
17&1&1504.59&0.12&0\\
17&2&1546.04&0.15&0\\
18&0&1551.38&0.17&0\\
18&1&1584.44&0.23&0\\
18&2&1625.84&0.37&0\\
19&0&1630.95&0.24&0\\
19&1&1664.70&0.29&0\\
19&2&1706.39&0.54&0\\
20&0&1711.43&0.28&0\\
20&1&1745.50&0.42&0\\
20&2&1786.63&1.53&0\\
21&0&1792.82&1.27&0\\
21&1&1825.31&0.85&0\\
\hline 
\end{tabular}
\tablefoot{Columns $n$ and $\ell$ report the radial order and the spherical degree of the extracted modes. The second and third columns, $\nu$ and $\sigma_\nu$ provide the extracted frequency and its maximum uncertainty. The final column provides the flag for certainty, whereby 0 is a bonafide mode and 1 indicates a possible detection. Only modes with a flag=0 were used in the stellar modeling.}
\label{tab:individualFrequencies}
\end{table}

\begin{figure}
    \centering
    \includegraphics[width=\columnwidth]{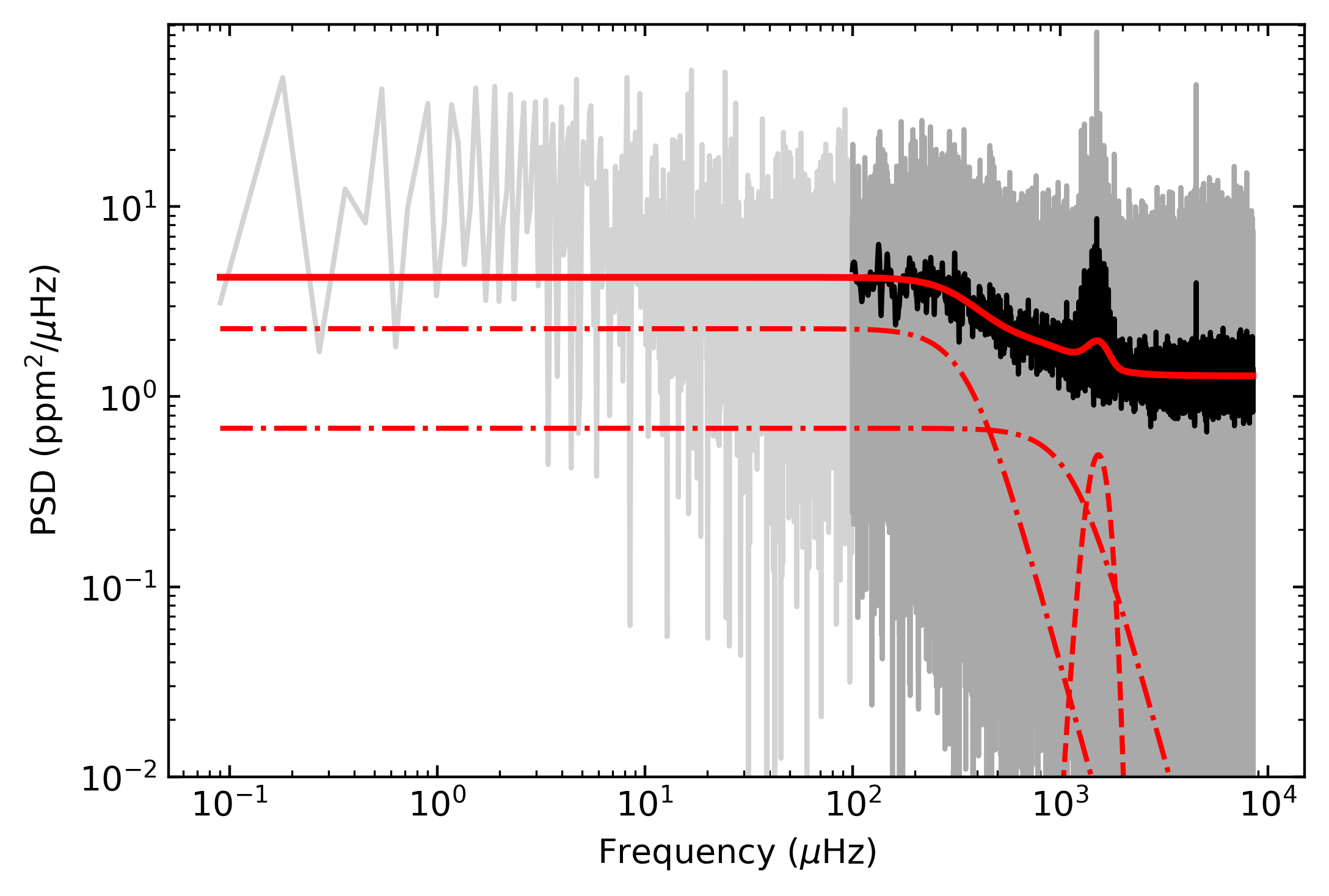}
    \caption{PSD of the primary of \KIC. The multi-component fit to the PSD and (upper panel) and the power excess (lower panel) are shown. In the upper panel,  the dash-dotted liens represent the background fits, while the dashed red line represents the gaussian fit. The solid line represents the combined fit to the PSD. 
    The vertical dashes in the lower panel indicate the of the extracted radial (blue), dipole (dark red), and quadruple (orange). 
    The solid red line depicts the combined solution of all~extracted~modes.}
    \label{fig:PSD}
    \label{fig:psdPattern}
\end{figure}

\subsection{Details on masses from asteroseismic scaling relations \label{sec:ScalingRelations}}

We calculated the masses and radii for the primary of \KIC, using the asteroseismic scaling relations, given in Eq.\,\ref{eq:mass} and \ref{eq:radius}, respectively. We use two different sets of global seismic parameters and solar reference values, in addition to T$_\mathrm{eff}$ from our spectroscopic analysis (see Table\,\ref{tab:spectroscopicParameters}).

The first set of \num and \dnu, as reported in the top panel of Table\,\ref{tab:globalSeismology}, has been determined by with the A2Z pipeline and was reported by \cite{Mathur2022}. Hereby, we use the solar reference values of the A2Z pipeline $\nu_{\mathrm{max},\odot}$\,=\,3100\,$\mu$Hz, $\Delta \nu_\odot$ = 135.2\,$\mu$Hz, and \hbox{T$_\mathrm{eff,\odot}$= 5777\,K \citep{Mathur2010}.} 

Because of the relatively large uncertainty of the power excess in the values obtained with A2Z, we reanalyzed this star using the universal pattern module of the \texttt{apollinaire}\protect\footnote{Documentation is available at \url{https://apollinaire.readthedocs.io/en/latest/}.} software package \citep{2022A&A...663A.118B}, which implements the \texttt{emcee} ensemble MCMC sampler \citep{2013PASP..125..306F}. The revised values for \num and \dnu are provided in the middle panel of Table\,\ref{tab:spectroscopicParameters}. We derived the solar reference values adopted by the \texttt{apollinaire} code from observations obtained with the Sun photometers (SPMs) of the Variability of solar IRradiance and Gravity Oscillations instrument \citep[VIRGO,][]{Froehlich1997} onboard the Solar and Heliospheric Observatory \citep[SoHO,][]{Domingo1995}. Using a one-year time series from the VIRGO/SPM green and red channels acquired at the beginning of the SoHO mission, we measured $\nu_{\mathrm{max},\odot}=3073.98 \pm 7.27,\mu\mathrm{Hz}$ and $\Delta\nu_\odot=134.86 \pm 0.02,\mu\mathrm{Hz}$.

\begin{figure*}[th!]
    \centering
    \includegraphics[width=\textwidth]{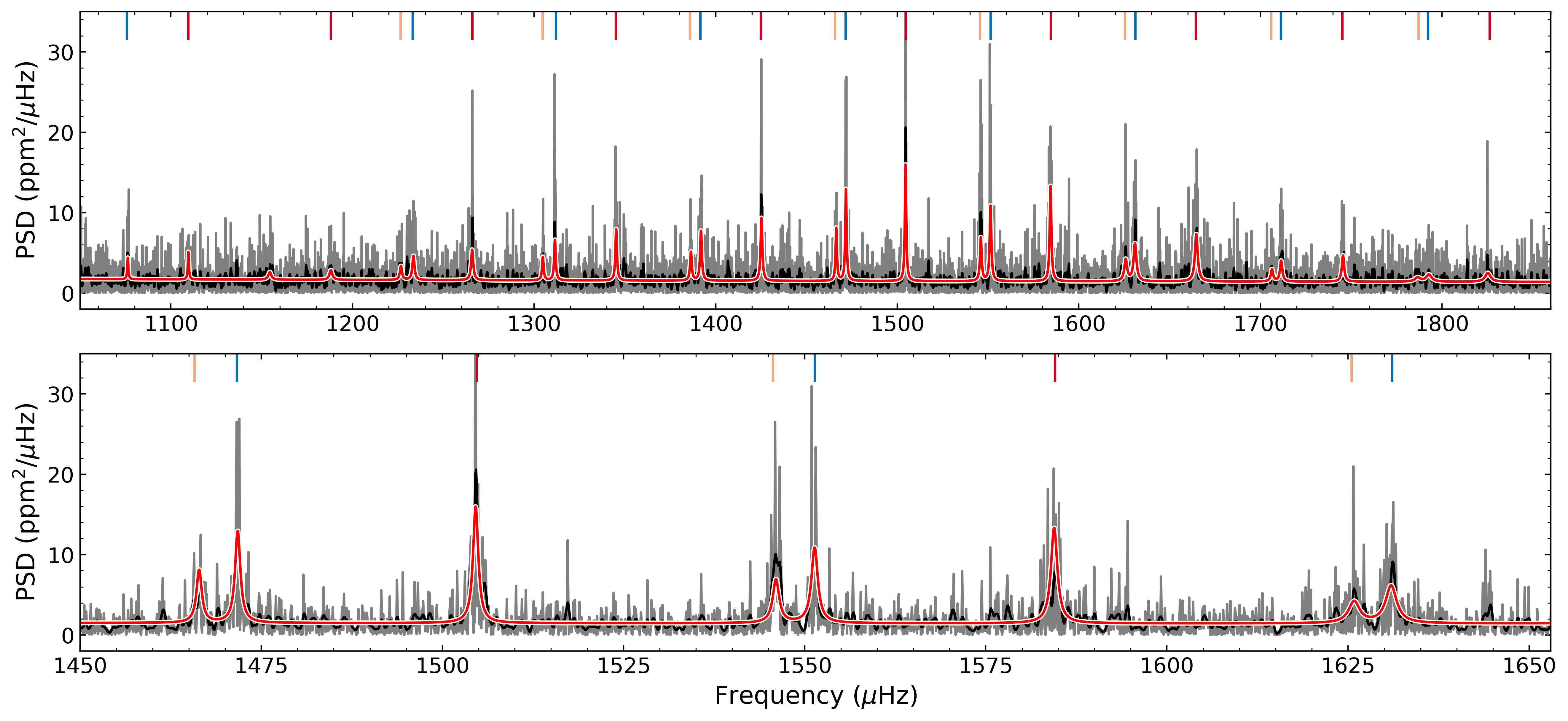}
    \caption{PSD of \KIC around the excess of power spectral density. The top panel shows the full power excess, with the original, and smoothed PSD in grey, and black, respectively. The lower provides a zoomed view on the two central radial orders.
    The red line depicts the combined model of extracted frequencies, which are represented by the vertical dashes (blue, red, and orange for $\ell$=0,\,1,\,and\, 2, resp., see Table\,\ref{tab:individualFrequencies}).}
    \label{fig:peakbaggedPSD}
\end{figure*}

In order to account for the deviation in stars from the homological asteroseismic scaling relations as given in Eq.\,\ref{eq:radius} and Eq.\,\ref{eq:mass}, which use the Sun as a scaling reference, and also accounting for effects resulting from poor modeling of the surface, in the past decade corrections have been introduced to recalibrate those relations \citep[][ and references therein]{Hekker2020rev, Li2022}.

One such correction, is provided by \cite{Mosser2013}, where in order to account for the deviation of the observed \dnu from the asymptotic frequency spacing $\Delta\nu_{\mathrm{as}}$, a term is introduced to correct the observed frequency spacing $\Delta\nu_{\mathrm{obs}}$ given as
\begin{equation}
    \Delta\nu_{\mathrm{as}} = (1 -\zeta)\,\Delta\nu_{\mathrm{obs}}, 
\end{equation}
where 
\begin{equation}
    \zeta = \frac{0.57}{n_{\mathrm{max}}},\, n_{\mathrm{max}} = \frac{\nu_{\mathrm{max}}}{\Delta\nu_{\mathrm{obs}}}.
\end{equation}

More recently, \cite{Li2023} introduced a correction, derived empirically from \Kepler RGB and SGB stars, that depends on atmospheric parameters ($T_\mathrm{eff}$ and $[$Fe/H$]$) and the global asteroseismic parameters \num and \dnu. 
\KIC is well within the parameter space provided by the grid to calculate the correction term.
Using those corrections on the target, results in higher mass and radius, and are given in Table\,\ref{sec:ScalingRelations} with the suffix L23.
For both sets of global seismic parameters, we test the seismic scaling relation in the uncorrected from as well as with the correction formalisms of \cite{Mosser2013} and \cite{Li2023}. The corresponding values are reported in Table\,\ref{tab:globalSeismology} and depicted in Fig.\,\ref{fig:scalingRelationComparison}.

\subsection{Peakbagging of individual modes \label{sec:peakbagging}}

To characterize the oscillation modes of KIC~9693187, we again use the \texttt{apollinaire} software package. 
Each mode was modeled as a Lorentzian profile, with its central frequency and linewidth treated as independent parameters. A single mode height was fitted collectively for all $\ell = 0, 1$ modes of order $n$ and for $\ell = 2$ modes of order $n-1$. Additionally, the relative height ratios between different angular degrees were included as global parameters in the fit.

For the sampling process, logarithmic forms of the amplitude and linewidth, as well as the frequency and relative mode heights, are used to ensure non-informative, uniform priors. All parameters of all modes are fitted at the same time in a global way \citep[as first done by][]{1999ESASP.448..135R}.

Posterior distributions are generated for all parameters, from which the median values are adopted as best estimates. The associated uncertainties are defined by the larger deviation between the median and the 16$^\mathrm{th}$ or 84$^\mathrm{th}$ percentiles of the posterior samples. For the final error bars given in Table\,\ref{tab:individualFrequencies}, and following a conservative approach, the larger of the asymmetrical error obtained during the sample of the posteriors is provided. The PSD, overplotted with the model of the extracted modes is depicted in Fig.\,\ref{fig:peakbaggedPSD}.
\begin{figure}[t!]
    \centering \includegraphics[width=\columnwidth,height=45mm]{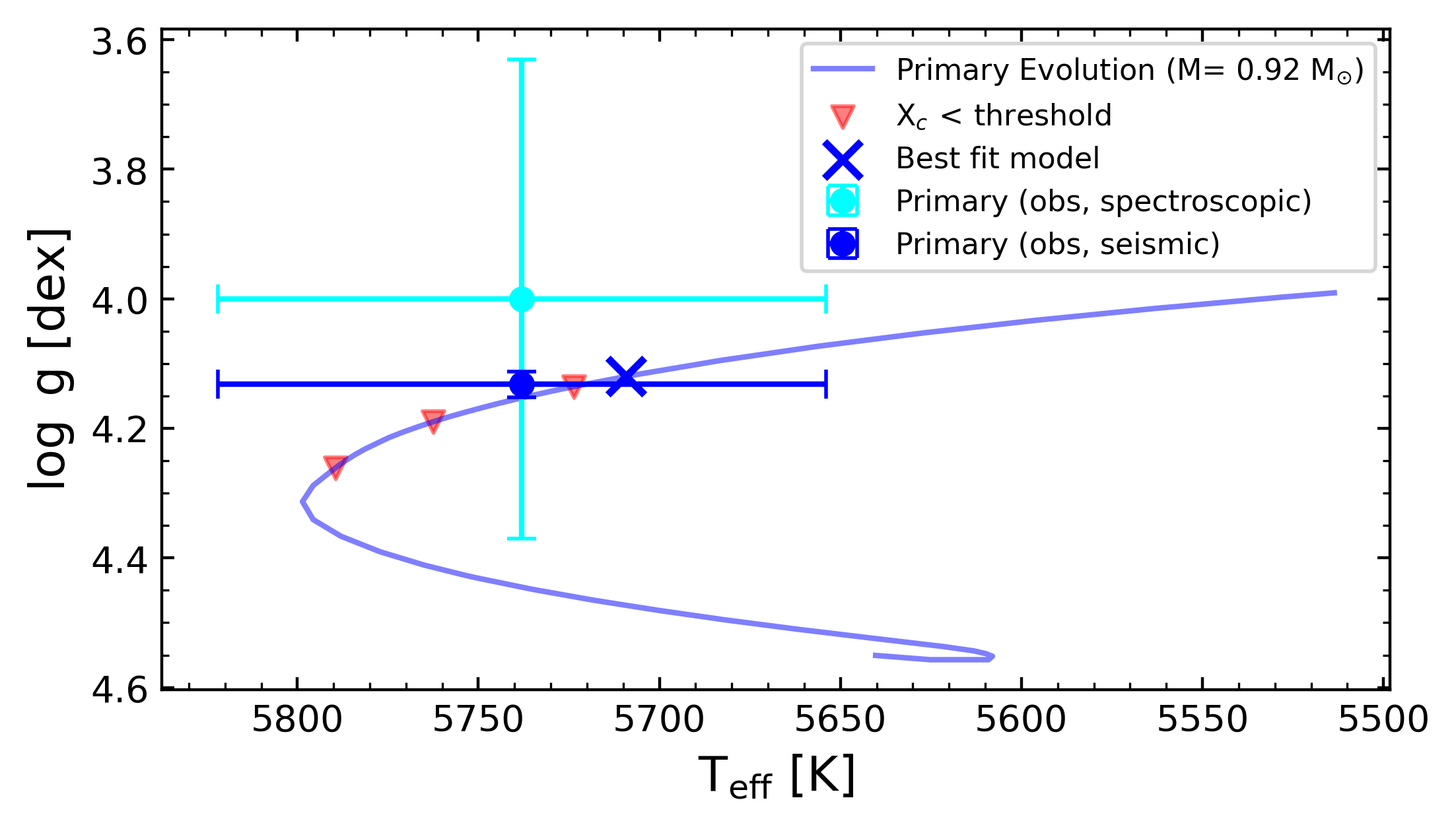}
    \vspace{-7mm}
    \caption{Kiel diagram of the primary of \KIC. The spectroscopic and seismic $\log g$ are depicted for the primary. The evolutionary track of the MS and the subgiant phase is shown. The red triangles show when the fractional core-H content X$_c$ drops below 1\%, 0.01\%, and 0.0001\% and the blue cross gives the position of the best fit model.
    }
    \label{fig:HRD}
\end{figure}

\subsection{Individual frequency modeling with MESA \label{sec:AppB_mesa}}

To estimate the mass, radius and age of the oscillating component, we performed a detailed seismic grid modeling of \KIC.
As input we used the effective temperature from the spectroscopic analysis of the LCO data. 
We corrected the metallicity following 
\cite{Salaris1993},
\begin{eqnarray}
    [\mathrm{M/H}] = [\mathrm{Fe/H}] + \log_{10}(0.638 \cdot 10^{\mathrm{[\alpha/Fe]}} +0.362),
\end{eqnarray}
to account for the modified mean molecular weight in the  the enrichment of $\alpha$-elements. As input metallicity, we used \hbox{[M/H]=-0.18\,$\pm$0.15dex}.

The search for the best fitting model was performed through an iterative analysis, using the IACgrid. This collection of pre-calculated evolutionary tracks for mains-sequence stars has been computed with the MESA \citep[Modules for Experiments in Stellar Astrophysics; version 15\,140;][]{MESA2023,Paxton2011} code package.
The models adopt standard input physics, utilizing OPAL opacity tables \citep{Iglesias1996} and the chemical composition described by \citep{Grevesse1998}. Stellar masses span from 0.8 to 1.5\,$M_\odot$ in increments of 0.01\,$M_\odot$, with initial metallicities ranging from -0.3 to +0.4 dex in steps of 0.05. The mixing-length parameter, $\alpha$, varies between 1.5 and 2.2, also in steps of 0.05, following the formulation by \cite{Cox1968}. Oscillation frequencies are computed using the adiabatic pulsation code ADIPLS \citep{JcD2008}, and the initial helium content is fixed at $Y_0$\,=\,0.249. 

The global seismic parameters (Table\,\ref{tab:globalSeismology}), and the individual frequencies of the modes from Table\,\ref{tab:individualFrequencies} with flag 0 were used. 
Model fitting is performed through $\chi^2$ minimization, where separate contributions from spectroscopic constraints, oscillation frequencies, and dynamical properties are considered.
The dynamical component accounts for the characteristic timescale $(R^3/GM)^{0.5}$. Surface corrections were applied following \cite{PerezHernandez2019}. Further methodological details, including the treatment of uncertainties, are provided in \cite{PerezHernandez2019} and \cite{GonzalezCuesta2023}.

While the $T_\mathrm{eff}$, metallicity,  are input parameters, the large number of individual frequencies dominates the error, leading to smaller uncertainties and robust values \citep{LebretonGoupil2014, Grossmann2025}. We also tested several additional constrains, such as the luminosity and surface gravity. However, these had limited impact on constraining the best fit model as these were outnumbered by the individual frequencies.

Based on the best fit model parameters, we calculated the evolutionary track for the primary, that is shown in Fig.\,\ref{fig:HRD}. We used the age of 12 Gyr, above 1-$\sigma$ of the models age, as a stopping criterion. 
The bottom panel of Table\,\ref{tab:globalSeismology} presents the mass (M$_\mathrm{IF}$), radius (R$_\mathrm{IF}$), and age (A$_\mathrm{IF}$) determined for \KIC from the individual-frequency modeling (IF).

Figure\,\ref{fig:HRD} shows the position of \KIC according to the spectroscopic and seismic surface gravity in blue and purple, respectively. Furthermore we mark the position of the best fit model from the individual frequency modeling in the HRD.
To test if \KIC's primary is located before or after the turnoff from the MS, we also marked the depletion of the fractional 
content of hydrogen in the core, whereby we mark the positions X$_{c}\,<\,1 \%,0.01\%$ and $0.0001\%$, confirming that the star has consumed almost its core hydrogen.

\end{appendix}
\label{LastPage}
\end{document}